\documentclass[aps,pra,twocolumn,superscriptaddress]{revtex4}

\usepackage{amsfonts,epsfig}
\usepackage{latexsym}
\usepackage{epsfig}
\usepackage{amsmath}
\usepackage{amssymb}
\usepackage{amsfonts}
\usepackage{amsthm}
\usepackage{mathrsfs}
\usepackage{natbib}
\usepackage{color,verbatim,graphics}
\usepackage{psfrag}
\DeclareMathAlphabet{\mathrsfs}{U}{rsfs}{m}{n}
\DeclareMathAlphabet{\mathpzc}{OT1}{pzc}{m}{it}
\DeclareMathAlphabet{\matheus}{U}{eus}{m}{n}
\DeclareMathAlphabet{\mathbbold}{U}{bbold}{m}{n}

\begin{comment}

\newcommand{\ket}[1]{\left | #1 \right \rangle}
\newcommand{\bra}[1]{\left \langle #1   \right |}

\newcommand{\comment}[1]{}

\newcommand{\ba}{\begin{eqnarray}}
\newcommand{\be}{\begin{equation}}
\newcommand{\ee}{\end{equation}}

\newcommand{\ea}{\end{eqnarray}}
\newcommand{\ban}{\begin{eqnarray*}}
\newcommand{\ean}{\end{eqnarray*}}

\newcommand{\ket}[1]{|#1\rangle}
\newcommand{\bra}[1]{\langle#1|}

\newcommand{\ketbra}[2]{|#1\rangle\langle#2|}

\begin{document}

\title{Long-distance entanglement generation with scalable and robust two-dimensional quantum network}

\author{Ying Li} 
\affiliation{Centre for Quantum Technologies, National University of Singapore, 2 Science Drive 3, Singapore 117543}
\author{Daniel Cavalcanti} 
\affiliation{Centre for Quantum Technologies, National University of Singapore, 2 Science Drive 3, Singapore 117543}
\author{Leong Chuan Kwek}
\affiliation{Centre for Quantum Technologies, National University of Singapore, 2 Science Drive 3, Singapore 117543}
\affiliation{National Institute of Education and Institute of Advanced Studies, Nanyang Technological University, 1 Nanyang Walk, Singapore}

\date{\today, version 5}

\begin{abstract}
We present a protocol for generating entanglement over long distances in a two-dimensional quantum network based on the surface-error-correction code.
This protocol requires a fixed number of quantum memories at each node of the network and tolerates error rates of up to $1.67\%$ in the quantum channels.
If local operations are of an error rate of $10^{-3}$, error rates of about $1.27\%$ in the quantum channels are tolerable.
\end{abstract}

\pacs{XXX}
\maketitle

\section{Introduction}

Creating entanglement over long distances is the main goal of quantum communication, with applications in quantum key distribution, fundamental tests of quantum mechanics, and distributed computing among others \cite{gisin-thew}.
However, the fragility of entanglement to environmental noise limits the effective distance of direct quantum communication.
One of the most celebrated solutions to this problem is the use of quantum repeaters \cite{repeater}.
As a drawback, this strategy consumes an amount of quantum memories per repeater that grows rapidly with the distance for establishing entanglement, even when error correction is used \cite{L.Jiang, surface}.

The distribution of entanglement in quantum networks has been the focus of intense research.
Nontrivial geometry of the quantum network can be used, for instance, in entanglement percolation \cite{ent_perc} or error-correction strategies \cite{raussendorf,perseguers2,perseguers3D,AGrudka}.
However, all the known results in this direction rely on unrealistic quantum states \cite{ent_perc,lapeyre1,perseguers1,multipartite,broadfoot1,broadfoot2,broadfoot3} or networks with impractical geometries (e.g. three dimensional) \cite{raussendorf,perseguers3D,AGrudka} or the consumption of a growing amount of local resources \cite{perseguers2,lapeyre2}.
Entanglement distribution in a noisy two-dimensional network with fixed local resources is believed to be possible through one-dimensional fault-tolerant quantum-computation schemes \cite{perseguers2,AGrudka}.
However, such a scheme often requires quantum communications and operations with a very small error rate (approximately $10^{-5}$) \cite{1dcode}.
Thus, the problem of designing a realistic scalable quantum network  remains largely unresolved.

In this paper, we show that it is possible to entangle two distant sites in a two-dimensional network involving realistic quantum channels.
In the present proposal, the number of quantum memories per node needed is fixed and does not scale with the communication distance.
Also, the scalability of the two-dimensional quantum network does not rely on the scalability of quantum processors.
Moreover, quantum-communication error rates of up to $1.67\%$ can be tolerated.

Our starting point is a quantum network on the square lattice (see Fig. \ref{scheme}). 
Each node in the network is connected to its neighbors through a quantum channel that distributes two-qubit Werner states $\rho$ given by
\ba\label{werner}
\rho=(1-q)\ketbra{\Phi_+}{\Phi_+}+q\frac{\openone}{4},
\ea
where $\ket{\Phi_+}=(\ket{00}+\ket{11})/\sqrt{2}$ is a maximally entangled state, $\openone/4$ is the maximally mixed state, and $0\leq q \leq 1$ is a noisy parameter. This state can be understood as the result of the following process: a maximally entangled state $\ket{\Phi_+}$ is produced and sent to a neighboring site through a depolarizing channel. This channel leaves the state untouched with probability $F=\bra{\Phi_+}\rho\ket{\Phi_+}=1-3q/4$ (i.e. the fidelity between $\rho$ and $\ket{\Phi_+}$) and causes an error with probability $1-F$, which we call the channel-error rate.   
Note that since any two-qubit state can be put into the form \eqref{werner} by local operations and classical communication \cite{bennett}, our results can also cover other cases of quantum states. 
The main goal in our scheme is to entangle two arbitrarily distant nodes, labeled by Alice and Bob, using quantum channels connecting neighboring nodes, local operations at each node and one-way classical communication among them.
Here, we will consider, apart from the communication noise, possible errors in these operations.  
Our protocol is based on the surface code \cite{surfacecode} and could be generalized to other geometries \cite{colorcode}.
Apart from the four qubits in each node composing the network, we need one more qubit in each node for processing the surface code. 

\begin{figure*}[tbp]
\includegraphics[width=15 cm]{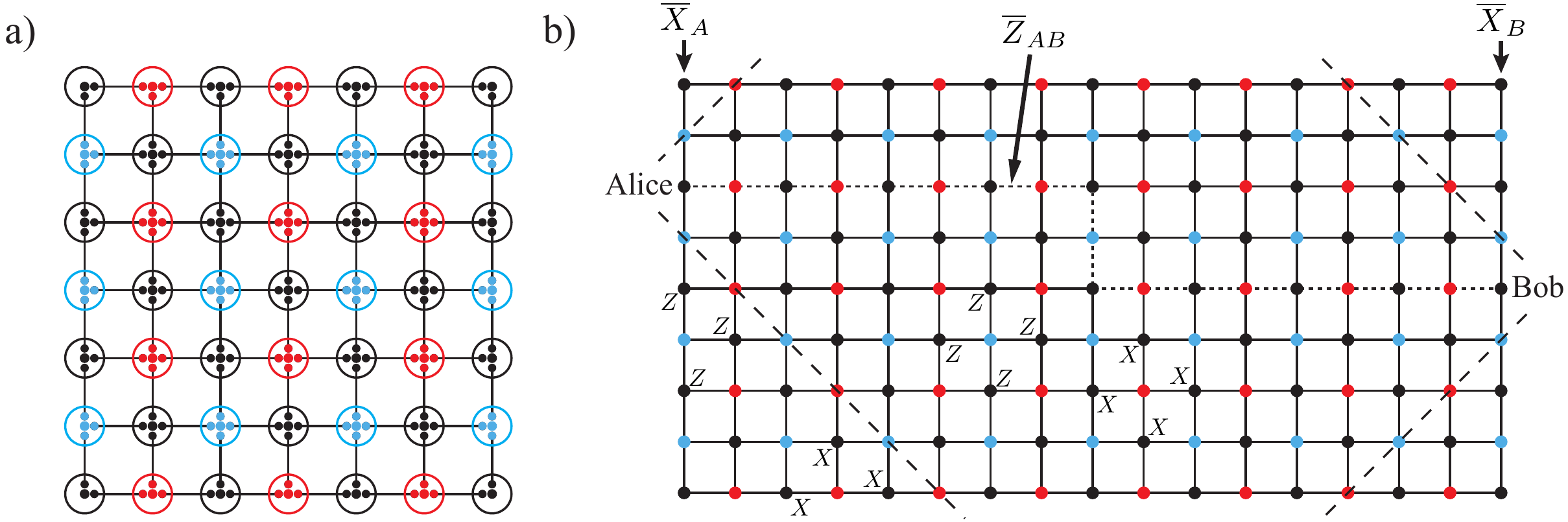}
\caption{
(a) Quantum network on the square lattice.
Each node has four linking qubits, which can be entangled with neighboring nodes, while the fifth one is used to process the surface code.
The colors are used to label the nodes according to the operations to be realized during the protocol.
(b) A rectangular part of the quantum network is used to create entanglement between qubits in Alice's and Bob's sites (see text).
}
\label{scheme}
\end{figure*}

\begin{figure}[h]
\includegraphics[width=8cm]{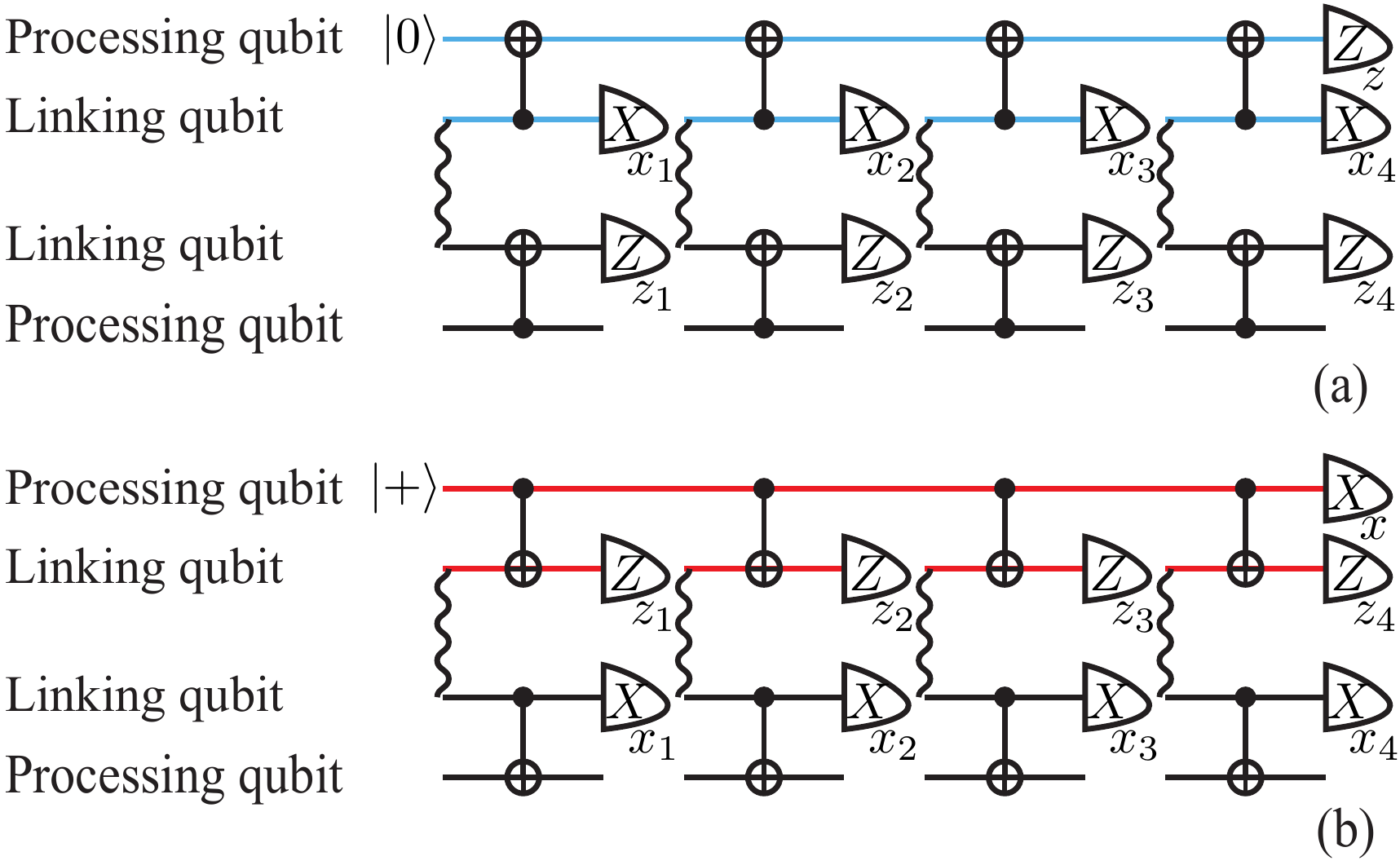}
\caption{
Circuits for stabilizer measurements (a) $ZZZZ$ and (b) $XXXX$.
Circuits for stabilizer measurements $ZZZ$ and $XXX$ are similar.
On each subfigure, the upper two lines are the processing qubit and linking qubits of a \textit{blue} or \textit{red} node, while the lower two lines are processing qubits and linking qubits of four neighboring \textit{black} nodes.
Each wave line represents a Bell state $\left\vert\Phi ^{+} \right\rangle$ of two corresponding linking qubits.
The measurement outcome of $ZZZZ$ ($XXXX$) is $zz_{1}z_{2}z_{3}z_{4}$ ($xx_{1}x_{2}x_{3}x_{4}$) where $z$ and $z_{i}$ ($x$ and $x_{i}$) are outcomes of measurements in the $Z$ ($X$) basis of the \textit{blue} (\textit{red}) processing qubit and the $i$th black linking qubit, respectively.
Each \textit{blue} (\textit{red}) node interacts with its four neighboring \textit{black} nodes in the order left, up, right, down.
After interacting with a \textit{blue} (\textit{red}) node, a \textit{black} processing qubit needs a phase (flip) gate $Z^{x_{i}}$ ($X^{z_{i}}$) where $x_{i}$ ($z_{i}$) is the measurement outcome of the corresponding \textit{blue} (\textit{red}) linking qubit.
}
\label{circuit}
\end{figure}

\section{Scheme}

To generate remote entanglement between Alice and Bob, we consider a section of the network with a rectangular geometry as shown in Fig. \ref{scheme}(b). We divide the nodes within this section of the network into three groups, marked in \textit{black}, \textit{blue}, and \textit{red} in the figure.
Each \textit{blue} (\textit{red}) node is surrounded by four \textit{black} nodes (or three if it is along a border of the rectangle).
Alice and Bob are both in the \textit{black} group and are located on two edges in this rectangular network, e.g. two vertical sides composed of \textit{black} nodes and \textit{blue} nodes.
The other two sides are composed of \textit{black} nodes and \textit{red} nodes.
At the start of the protocol, we initialize all processing qubits in \textit{black} nodes to the state $|0\rangle$.
We then use the entanglements shared between neighbors to perform stabilizer measurements $ZZZZ$ ($ZZZ$) and $XXXX$ ($XXX$) of four (three) \textit{black} processing qubits around each \textit{blue} and \textit{red} node, respectively.
Here, $Z$ and $X$ are Pauli operators.
A circuit describing these measurements is shown in Fig. \ref{circuit}.
As soon as these stabilizer measurements are performed, the state of \textit{black} processing qubits becomes an eigenstate of the stabilizers of the surface code \cite{surfacecode}.
Finally, all \textit{black} processing qubits except Alice's and Bob's qubits are measured in the following way:
\begin{itemize}
\item All \textit{black} processing qubits along the two vertical sides are measured in the $X$ basis.
\item All \textit{black} processing qubits along the dotted line composed of \textit{black} and \textit{red} nodes connecting Alice and Bob [see Fig. \ref{scheme}(b)] are measured in the $Z$ basis.
\item Qubits in the region defined within the dashed lines in Fig. \ref{scheme}(b) are measured in the $Z$ basis, and the ones outside are measured in the $X$ basis.
\end{itemize}
Here, we choose the dotted line so that it is in the middle of two corresponding dashed lines when it is near Alice and Bob or two horizontal lines when it is far away from Alice and Bob.
We argue that after these measurements, the processing qubits of Alice and Bob are  entangled.

In order to see this, let us first consider the perfect case, i.e. when $q=0$ and all operations are perfect.
The initial state of \textit{black} processing qubits, which are all initialized in the state $|0\rangle$, is the eigenstate of $\overline{Z}_{AB}$ with the eigenvalue $+1$.
Here, $\overline{Z}_{AB}$ is the product $\prod Z$ of \textit{black} processing qubits on the line connecting Alice and Bob (the dotted line in Fig. \ref{scheme}).
The operator $\overline{Z}_{AB}$ commutes with the stabilizer operators.
Therefore, the stabilizer state is still an eigenstate of $\overline{Z}_{AB}$ with the eigenvalue $+1$.
The stabilizer state is also an eigenstate of the product of all $XXXX$ and $XXX$, which is $\overline{X}_{A}\overline{X}_{B}$, where $\overline{X}_{A}$ ($\overline{X}_{B}$) is the product $\prod X$ of \textit{black} processing qubits on the vertical side with Alice (Bob) [see Fig. \ref{scheme}(b)].
One can obtain the eigenvalue of $\overline{X}_{A}\overline{X}_{B}$ by multiplying measurement outcomes of all $XXXX$ and $XXX$.
After measuring out \textit{black} processing qubits except the processing qubits in Alice and Bob (i.e., the qubit $A$ and the qubit $B$), we can replace $Z$ and $X$ in $\overline{Z}_{AB}$ and $\overline{X}_{A}\overline{X}_{B}$ with the respective measurement outcomes.
Finally, we see that the state of qubits $A$ and $B$ is ``stabilized", i.e., it becomes an eigenstate of $Z_{A}Z_{B}$ and $X_{A}X_{B}$, where eigenvalues depend on measurement outcomes.
In this way, the qubit $A$ and the qubit $B$ are entangled as one of the Bell states.

Imperfections in quantum channels and in local operations can result in incorrect stabilizer-measurement outcomes.
In order to obtain a set of faithful stabilizer-measurement outcomes, the stabilizer measurements must be repeated $N$ times before final single-qubit measurements on \textit{black} processing qubits.
For each stabilizer measurement, the entanglement between neighboring sites needs to be regenerated. Thus, the overall time cost of our scheme is $NT$, where $T$ is the communication time for generating neighboring entanglements.

It is crucial to realize that \textit{black} processing qubits may be affected by errors during the stabilizer measurements.
However, these errors can be detected: if any of the stabilizer-measurement outcomes are different from each other in the previous time step, we have an error syndrome and we immediately conclude that incorrect stabilizer-measurement outcomes and errors on \textit{black} processing qubits have happened.
Moreover, it is possible that some qubits are wrongly initialized states other than the state $|0\rangle$ at the very beginning.
We can detect such initialization errors based on measurement outcomes of $ZZZZ$ stabilizers, i.e., all $ZZZZ$ should be $+1$ if the qubits are initialized correctly.
Errors occurring after the last stabilizer measurement, including errors induced by the last stabilizer measurement and subsequent operations, cannot be detected by further stabilizer measurements.
Thus, we may need to measure more \textit{black} processing qubits rather than only qubits included in $\overline{Z}_{AB}$ and $\overline{X}_{A}\overline{X}_{B}$ (see the measurement pattern defined by the dashed lines in Fig. \ref{scheme}).
We then detect these errors that occur after the last stabilizer measurements through a comparison of the outcomes of single-qubit measurements with outcomes of stabilizers, i.e., the outcome of a stabilizer should be the same as the product of outcomes of individual qubits in the stabilizer.
One corrects stabilizer-measurement outcomes and all other errors by pairing error syndromes \cite{SCQC} as in the typical surface-code error correction.

\begin{figure}[tbp]
\includegraphics[width=8 cm]{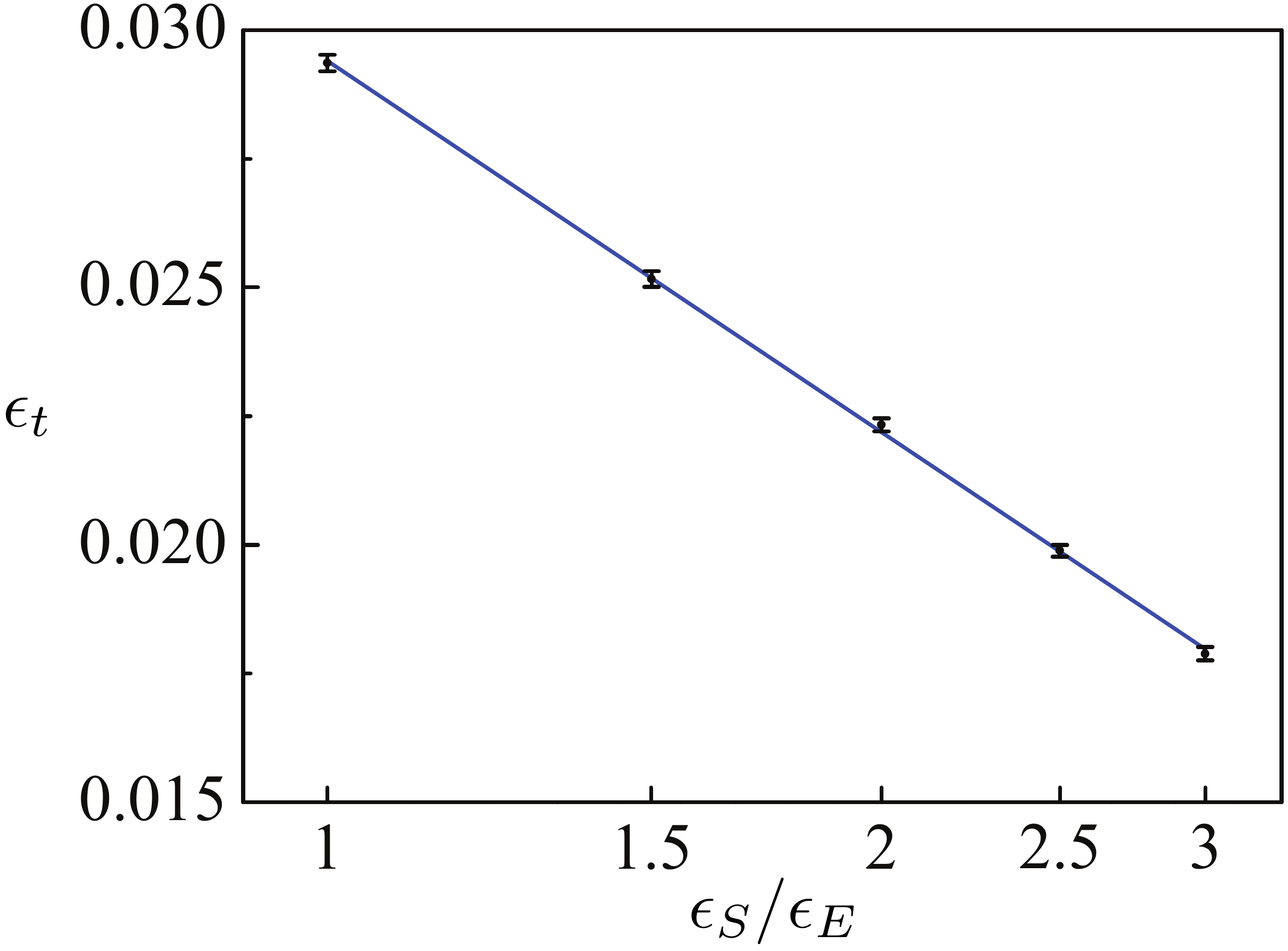}
\caption{
Error thresholds for a variety of ratios between $\epsilon_{S}$ and $\epsilon_{E}$ for independent errors.
Here, $\epsilon_{t}$ is the threshold of $\epsilon_{E}$, i.e., errors are correctable if $\epsilon_{E}<\epsilon_{t}$.
Squares represent thresholds for $\epsilon_{S}/\epsilon_{E}=1,1.5,2,2.5,3$ without correlations.
These thresholds are obtained numerically by pairing error syndromes with the minimum-weight perfect-matching algorithm \cite{MWPM,Sean}.
The line is obtained by fitting thresholds with the function $\epsilon_{t}=\epsilon_{0}-k\log(\epsilon_{S}/\epsilon_{E})$.
}
\label{thresholds}
\end{figure}

\section{Error thresholds}

The surface code works if the probability of errors is lower than a certain threshold.
The outcome of a $XXXX$ or $ZZZZ$ measurement may be wrong with a probability $\epsilon_{S}$.
Between two time steps of stabilizer measurements, phase errors $[Z]$ (flip errors $[X]$) may happen on each \textit{black} processing qubit with a probability $\epsilon_{E}$.
Here and throughout we use the form $[U]$ to denote the superoperator
$[U]( \rho ) = U\rho U^{\dagger}$.
By considering only the errors coming from quantum channels that occur ``independently" and by considering the limit where $q$ is small, $\epsilon_{S}=2q$, and $\epsilon_{E}=q$.
In fact, errors corresponding to $XXXX$ and $ZZZZ$ stabilizers are correlated.
However, these two kinds of errors can be corrected separately.
Thus, correlations between them can be ignored.
Under these conditions, we find numerically that the error threshold depends on the ratio $\epsilon_{S}/\epsilon_{E}$ as $\epsilon_{t}=\epsilon_{0}-k\log(\epsilon_{S}/\epsilon_{E})$ (see Fig. \ref{thresholds}), where $\epsilon_{0}=0.0294$ and $k=0.0072$ are constants and $\epsilon_{th}$ is the threshold of $\epsilon_{E}$ (i.e. errors are correctable if $\epsilon_{E}<\epsilon_{th}$).
In our case in which $\epsilon_{S}/\epsilon_{E}=2$, the noise in quantum channels is correctable if $q=\epsilon_{E}<2.23\%$, corresponding to an error rate of $1.67\%$.

Imperfect operations, including the initialization of qubits, measurements, and controlled-NOT gates, may also result in errors, reducing the tolerable error rate of quantum channels.
Without loss of generality, we may assume that errors in operations are depolarized with the same rate $p$.
Erroneous operations are modelled by perfect operations preceded or followed by an erroneous superoperation $E_{1}=(1-p)[I]+(1/3)([X]+[Y]+[Z])$ for single-qubit operations or 
$E_{2}=(1-p)[I]+(1/15)([I_{1}X_{2}]+\cdots +[Z_{1}I_{2}]+\cdots +[X_{1}Y_{2}]+\cdots )$ for two-qubit operations.
Moreover, imperfect two-qubit gates may give rise to correlations between phase errors on \textit{black} processing qubits, which take place in the form $[Z_{red}Z_{right}]$, $[Z_{red}Z_{down}]$, and $[Z_{righ}Z_{down}]$ with the same probability $\epsilon_{C}$ between two time steps of stabilizer measurements.
Here, $[Z_{red}]$ is a phase error on a \textit{red} processing qubit, which can induce an incorrect outcome of the stabilizer measurement, and $[Z_{right}]$ ($[Z_{down}]$) is a phase error on the \textit{black} processing qubit to the right (downward direction) of the \textit{red} processing qubit.
All other phase errors are independent, i.e. $[Z_{red}]$, $[Z_{right}]$, and $[Z_{down}]$ happen with the probabilities $\epsilon_{S}-2\epsilon_{C}$, $\epsilon_{E}-2\epsilon_{C}$, and $\epsilon_{E}-2\epsilon_{C}$, respectively.
Flip errors corresponding to stabilizers $ZZZZ$ are also similar.
By counting these errors, we find $\epsilon_{S}=2q+124p/15$, $\epsilon_{E}=q+76p/15$, and $\epsilon_{C}=8p/15$.
Then, we evaluate the thresholds of quantum channels with imperfect operations as shown in Fig. \ref{ERT} and show that if the error rate of operations is $10^{-3}$, the threshold of $q$ is about $1.69\%$, corresponding to an error rate of $1.27\%$.

Memory errors can occur in our scheme while we are generating neighboring entanglements.
Fortunately, these memory errors can also be detected by stabilizer measurements, and the decoherence time does not have to be comparable to the overall time cost $NT$ but does for the communication time for generating neighboring entanglements $T$.
We suppose memory errors are given by depolarization and occur with the rate $p_{m}$ during the time $T$, which can increase $\epsilon_{E}$ by $2p_m/3$.
Thus, memory errors on processing qubits can lower the threshold but not dramatically with $p_{m}=10^{-2}$ as shown in Fig. \ref{ERT}.

\begin{figure}[tbp]
\includegraphics[width=8 cm]{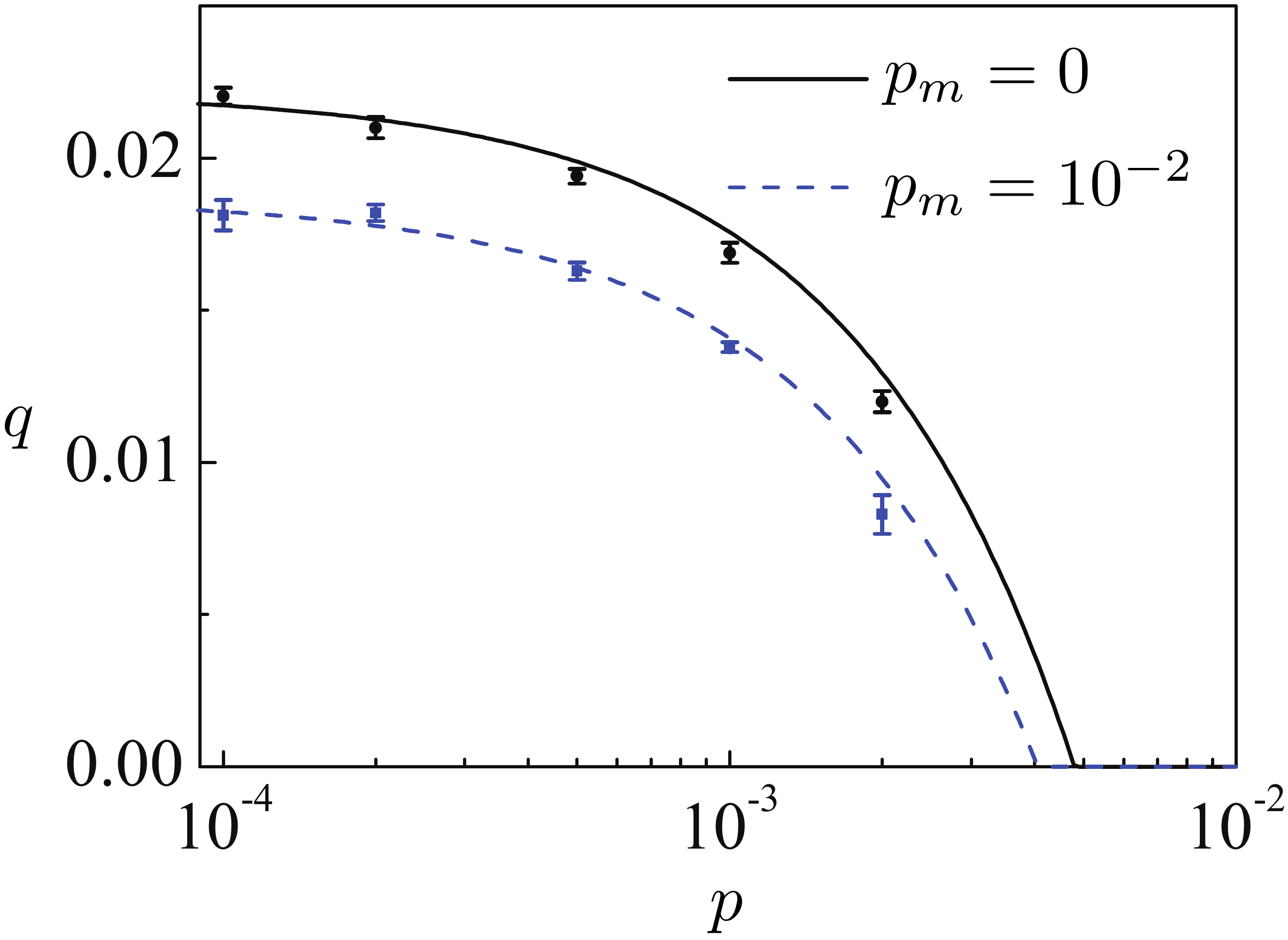}
\caption{
Thresholds of the communication-noise parameter $q$ where $p$ is the error rate of local operations.
Memory errors can lower the threshold but not dramatically with the error rate $p_{m}=10^{-2}$.
Two curves are obtained by neglecting correlations and using the linear fitting in Fig. \ref{thresholds}, which are good approximations of thresholds (rounds and squares for $p_{m}=0$ and $p_{m}=10^{-2}$, respectively) obtained numerically by pairing error syndromes with the minimum-weight perfect-matching algorithm \cite{MWPM,Sean}.
}
\label{ERT}
\end{figure}

\section{Final remote entangled state}

Even within the threshold, error correction may fail because a chain of errors connecting boundaries (error chain) may not be detected through error syndromes.
There are two kinds of nontrivial error chains that can affect the final entanglement between Alice and Bob: (i) error chains that flip qubits in $\overline{Z}_{AB}$ for an odd number of times and (ii) error chains that result in an odd total number of incorrect measurement outcomes of $XXXX$ stabilizers and phase errors on \textit{black} processing qubits along two vertical sides after the last stabilizer measurement.
In order to reduce the first kind of nontrivial error chains, the network for entangling Alice and Bob is designed so that the minimum distance between two horizontal sides and the line connecting Alice and Bob (the dotted line in Fig. \ref{scheme}) is also $N$.
Upon error correction, the total probability of long nontrivial error chains with the minimum length $N$ decreases exponentially with $N$ but increases polynomially with the distance between Alice and Bob \cite{raussendorf}.
Therefore, $N$ scales only logarithmically with the communication distance, and thus, these long error chains can then be neglected.
Short nontrivial error chains with lengths shorter than $N$ are all distributed in regions around Alice and Bob, whose probabilities also decrease exponentially with their lengths.
More generally, noise in the final remote entanglement can be described by the superoperator $E_{AB}=F[1]+\epsilon_{X}[X_A]+\epsilon_{Y}[Y_A]+\epsilon_{Z}[Z_A]$, where the fidelity $F=1-\epsilon_{X}-\epsilon_{Y}-\epsilon_{Z}$.
Assuming the last stabilizer measurement is $XXXX$, by only considering short error chains, we have $\epsilon_{X}=q/2+2p_{m}/3+44p/15+O(q^2,p_{m}^2,p^2)$, $\epsilon_{Y}=4p/15+O(q^2,p_{m}^2,p^2)$, and $\epsilon_{Z}=4p/3+O(q^2,p_{m}^2,p^2)$ \cite{raussendorf2,perseguers3D}.

\section{Efficiency}

The communication time for generating neighboring entanglements $T$ relies on the distance between two nearest-neighbor nodes.
For example, for nearest-neighbor distance of $10$~km, a neighboring entanglement can be generated with a probability $\sim 99.75\%$ in $T\sim 0.2$~ms.
Here, we have supposed a repeat of the entanglement generation in order to reach this high success probability, and the failure of generating entanglements is due to photon loss in fibers, whose attenuation is supposed to be $0.2$~dB/km in this example.
Failures of generating entanglements give rise to failures of stabilizer measurements, which are tolerable in surface codes.
The presence of these failures can reduce the threshold of noises, but only slightly if the success probability is near 1 \cite{Sean}.
With $L$ the distance between two vertical lines and $\sim N$ the distance between two horizontal lines, the probability of errors induced by long error chains scales as $\epsilon_{\text{long}}\sim Le^{-\kappa N}$. Here, $\kappa$ depends on the probability of errors, and $\kappa \sim 1$ for the probability of errors that is one-third of the threshold \cite{raussendorf3}.
Therefore, entanglements can be generated rapidly over a long distance, e.g. with $N=25$, resulting in $200$~ebits/s, and $\epsilon_{\text{long}}$ is still negligible if $L>10^5$, i.e. an overall distance of $10^6$~km.

\section{Discussion and conclusion}

In summary, we have proposed a protocol for entangling remote qubits on a two-dimensional noisy quantum network, which is scalable when the number of quantum memories in each node of the network is fixed.
In our protocol, the communication rate decreases only logarithmically with the distance.
The tolerable errors in the protocol presented here are three orders of magnitude better than possible protocols based on one-dimensional fault-tolerant quantum-computation schemes.
In this paper, we investigated the case in which each node has a five-qubit quantum memory.
Because every node is interacting with only one other node at one time, memory qubits can be reused, and indeed, two qubits per node is sufficient.
With more memories, entanglement distillation protocols can be used to improve the effective fidelity of quantum channels \cite{distillation}, i.e. increase the error-rate threshold.

This work is supported by the National Research Foundation and the Ministry of Education of Singapore.
D.C. acknowledges the PVE-CAPES program (Brazil). Y.L. acknowledges helpful discussions with Sean Barrett.


\begin{thebibliography}{99}

\bibitem{gisin-thew}N. Gisin and R. Thew, Nat. Photonics {\bf1}, 165 (2007) 

\bibitem{repeater} H. J. Briegel, W. D\"ur, J. I. Cirac, and P. Zoller, Phys. Rev. Lett. \textbf{81}, 5932 (1998).

\bibitem{L.Jiang} L. Jiang, J. M. Taylor, Kae Nemoto, W. J. Munro, Rodney Van Meter, and M. D. Lukin, Phys. Rev. A. \textbf{79}, 032325 (2009).

\bibitem{surface} A. G. Fowler, D. S. Wang, C. D. Hill, Thaddeus D. Ladd, R. Van Meter, and Lloyd C. L. Hollenberg, Phys. Rev. Lett. \textbf{104}, 180503 (2010).

\bibitem{ent_perc} A. Ac\'\i n, I Cirac, and M. Lewenstein, Nat. Phys. \textbf{3}, 256 (2007).

\bibitem{raussendorf} R. Raussendorf, S. Bravyi, and J. Harrington, Phys. Rev. A \textbf{71}, 062313 (2005).

\bibitem{perseguers2} S. Perseguers, L. Jiang, N. Schuch, F. Verstraete, M. D. Lukin, J. I. Cirac, and K. G. H. Vollbrecht, Phys. Rev. A \textbf{78}, 062324 (2008).

\bibitem{perseguers3D} S. Perseguers, Phys. Rev. A \textbf{81}, 012310 (2010).

\bibitem{AGrudka} A. Grudka, M. Horodecki, P. Horodecki, P. Mazurek, L. Pankowski, and A. Przysiezna, arXiv:1202.1016.

\bibitem{perseguers1} S. Perseguers, J. I. Cirac, A. Ac\'in, M. Lewenstein, and J. Wehr, Phys. Rev. A \textbf{77}, 022308 (2008).

\bibitem{lapeyre1} G. J. Lapeyre, J. Wehr, and M. Lewenstein, Physical Review A \textbf{79}, 042324 (2009).

\bibitem{multipartite} S. Perseguers, D. Cavalcanti, G. J. Lapeyre Jr, M. Lewenstein, and A. Ac\'n, Phys. Rev. A \textbf{81}, 032327 (2010).

\bibitem{broadfoot1} S. Broadfoot, U. Dorner, and D. Jaksch, EuroPhys. Lett. \textbf{88}, 50002 (2009).

\bibitem{broadfoot2} S. Broadfoot, U. Dorner, and D. Jaksch, Phys. Rev. A \textbf{81}, 042316 (2010).

\bibitem{broadfoot3} S. Broadfoot, U. Dorner, and D. Jaksch, arXiv:1008.3584.

\bibitem{lapeyre2} G. J. Lapeyre Jr., S. Perseguers, M. Lewenstein, and A. Ac\'in, arXiv:1108.5833.

\bibitem{1dcode} A. M. Stephens, A. G. Fowler, and L. C. L. Hollenberg, Quantum Inf. Comput. 8, 330 (2008􏰎); A. M. Stephens and Z. W. E. Evans, Phys. Rev. A 80, 022313 (2009).

\bibitem{bennett} C. H. Bennett, D. P. DiVincenzo, J. A. Smolin, and W. K. Wootters, Phys. Revi. A {\bf54}, 3824 (1996).

\bibitem{surfacecode} A. Y. Kitaev, Ann. Phys. \textbf{303}, 2 (2003).

\bibitem{colorcode} H. Bombin and M. A. Martin-Delgado, Phys. Rev. Lett. \textbf{97}, 180501 (2006); \textbf{98}, 160502 (2007).

\bibitem{SCQC} A. G. Fowler, A. M. Stephens, and P. Groszkowski, Phys. Rev. A \textbf{80}, 052312 (2009).

\bibitem{MWPM} C. Wang, J. Harrington, and J. Preskill, Ann. Phys. (NY) \textbf{303}, 31 (2003).

\bibitem{Sean} S. D. Barrett and T. M. Stace, Phys. Rev. Lett. \textbf{105}, 200502 (2010).

\bibitem{raussendorf2} R. Raussendorf , J. Harrington, K. Goyal, Ann. Phys. \textbf{321}, 2242 (2006).

\bibitem{raussendorf3} R. Raussendorf , J. Harrington, K. Goyal, New J. Phys. \textbf{9}, 199 (2007).

\bibitem{distillation} D. Deutsch, A. Ekert, R. Jozsa, C. Macchiavello, S. Popescu, and A. Sanpera, Phys. Rev. Lett. \textbf{77}, 2818 (1996).

\end{thebibliography}
\end{document}